\documentclass[sigconf]{acmart}

\usepackage{cleveref}
\usepackage{booktabs}
\usepackage{multirow}
\usepackage[normalem]{ulem}
\useunder{\uline}{\ul}{}
\usepackage[table,xcdraw]{xcolor}

\usepackage{graphicx}
\usepackage{subcaption}     

\usepackage{amsmath}
\usepackage{amsfonts}
\usepackage{amssymb}

\usepackage{balance}
\usepackage{amsthm}
\usepackage{enumitem}
\usepackage{color}

\usepackage{multirow}

\usepackage{bbding}

\usepackage{hyperref}
\usepackage{verbatim}

\AtBeginDocument{%
  }

\copyrightyear{2026}
\acmYear{2026}
\setcopyright{cc}
\setcctype{by}
\acmConference[SIGIR '26] {Proceedings of the 49th International ACM SIGIR Conference on Research and Development in Information Retrieval}{July 20--24, 2026}{Melbourne, VIC, Australia.}
\acmBooktitle{Proceedings of the 49th International ACM SIGIR Conference on Research and Development in Information Retrieval (SIGIR '26), July 20--24, 2026, Melbourne, VIC, Australia}
\acmISBN{979-8-4007-2599-9/2026/07}
\acmDOI{10.1145/3805712.3809701}

\newcommand{\ie}{\emph{i.e., }}
\newcommand{\eg}{\emph{e.g., }}

\newcommand{\cf}{\emph{cf. }}

\acmSubmissionID{fp122}

\begin{document}


\title{SpecTran: Spectral-Aware Transformer-based Adapter for LLM-Enhanced Sequential Recommendation }



\author{Yu Cui}
\authornotemark[2]
\authornotemark[3]
\orcid{0009-0001-6203-3022}
\affiliation{%
  \institution{Zhejiang University}
  \city{Hangzhou}
  \country{China}}
\email{cuiyu23@zju.edu.cn}

\author{Feng Liu}
\orcid{0009-0004-9265-9431}
\affiliation{%
  \institution{OPPO Research Institute}
  \city{Shenzhen}
  \country{China}}
\email{liufeng4hit@gmail.com}

\author{Zhaoxiang Wang}
\orcid{0009-0007-4925-7703}
\affiliation{%
  \institution{OPPO Research Institute}
  \city{Shenzhen}
  \country{China}}
\email{steven.wangzx@gmail.com}

\author{Changwang Zhang}
\orcid{0009-0004-4193-7833}
\affiliation{%
  \institution{OPPO Research Institute}
  \city{Shenzhen}
  \country{China}}
\email{changwangzhang@foxmail.com}

\author{Jun Wang}
\orcid{0000-0002-0481-5341}
\affiliation{%
  \institution{OPPO Research Institute}
  \city{Shenzhen}
  \country{China}}
\email{junwang.lu@gmail.com}

\author{Can Wang}
\authornotemark[2]
\authornotemark[4]
\orcid{0000-0002-5890-4307}
\affiliation{%
  \institution{Zhejiang University}
  \city{Hangzhou}
  \country{China}}
\email{wcan@zju.edu.cn}

\author{Jiawei Chen}
\authornote{Corresponding author.}
\authornote{State Key Laboratory of Blockchain and Data Security, Zhejiang University.}
\authornote{College of Computer Science and Technology, Zhejiang University.}
\authornote{Hangzhou High-Tech Zone (Binjiang) Institute of Blockchain and Data Security.}
\orcid{0000-0002-4752-2629}
\affiliation{%
  \institution{Zhejiang University}
  \city{Hangzhou}
  \country{China}}
\email{sleepyhunt@zju.edu.cn}

\renewcommand{\shortauthors}{Cui et al.}

\begin{abstract}
Traditional sequential recommendation (SR) models learn low-dimensional item ID embeddings from user-item interactions,  often overlooking textual information such as item titles or descriptions. Recent advances in Large Language Models (LLMs) have inspired a surge of research that encodes item textual information with high-dimensional semantic embeddings, and designs transformation methods to inject such embeddings into SR models. These embedding transformation strategies can be categorized into two types, both of which exhibits notable drawbacks: 1) adapter-based methods suffer from pronounced dimension collapse, concentrating information into a few dominant dimensions; 2) SVD-based methods are rigid and manual, considering only a few principal spectral components while discarding rich information in the remaining spectrum. 

To address these limitations, we propose SpecTran, a spectral-aware transformer-based adapter that operates in the spectral domain, attending to the full spectrum to select and aggregates informative components.  A learnable spectral-position encoding injects singular-value cues as an inductive bias, guiding attention toward salient spectral components and promoting diversity across embedding dimensions. Across four real-world datasets and three SR backbones, it consistently outperforms strong baselines, achieving an average improvement of 9.17\%.


\end{abstract}




\maketitle

\section{Introduction}

Sequential recommendation (SR) is a canonical task in recommender systems that predicts the next item a user will engage with based on the user’s historical interaction sequence~\cite{kang2018self,hidasi2015session,xie2022contrastive,yang2023generic}. A central objective is to learn high-quality item embeddings that capture item characteristics and thus inform users' behaviors and preferences~\cite{li2020time,tang2018personalized}.  Conventional SR methods assign each item a unique ID and learn its embedding from user–item interactions, aiming to capture collaborative signals inherent in historical data~\cite{kang2018self,sun2019bert4rec}. While effective, these approaches rely solely on interactions and largely ignore rich textual side information (e.g., titles, descriptions),  which explicitly conveys item properties and can substantially improve recommendation.

Motivated by the strong semantic understanding of large language models (LLMs), recent work has explored leveraging LLMs to enrich item embeddings with semantic knowledge~\cite{achiam2023gpt,dubey2024llama}. The prevailing approach encodes item text into language embeddings via LLMs, and then fuse these with ID embeddings to combine collaborative signals with semantic knowledge~\cite{ren2024enhancing,ren2024representation,sun2024large}. A fundamental challenge stems from the intrinsic difference between language-embedding spaces and item-embedding spaces: Language embeddings are high-dimensional (\eg 4096) representations of semantic content, whereas item embeddings are typically lower-dimensional (\eg 64) vectors tailored to capture collaborative signals. This mismatch complicates effective fusion.

Two main families of approaches have emerged to bridge these spaces: (i) \textbf{Adapter‑based methods}~\cite{yuan2023go,hou2022towards,ren2024representation,liu2024llm}, which learn a parametric MLP adapter to map high‑dimensional language embeddings into the low‑dimensional item space under the SR objective; and (ii) \textbf{SVD‑based methods}~\cite{zhang2024id,zhang2025llminit,hu2025alphafuse}, which apply singular value decomposition (SVD) to language embeddings and retain principal spectral components (those with large singular values) for dimension alignment.

\begin{figure}[t]
    \centering 
    \begin{subfigure}{0.235\textwidth}
        \centering
        \includegraphics[width=\textwidth]{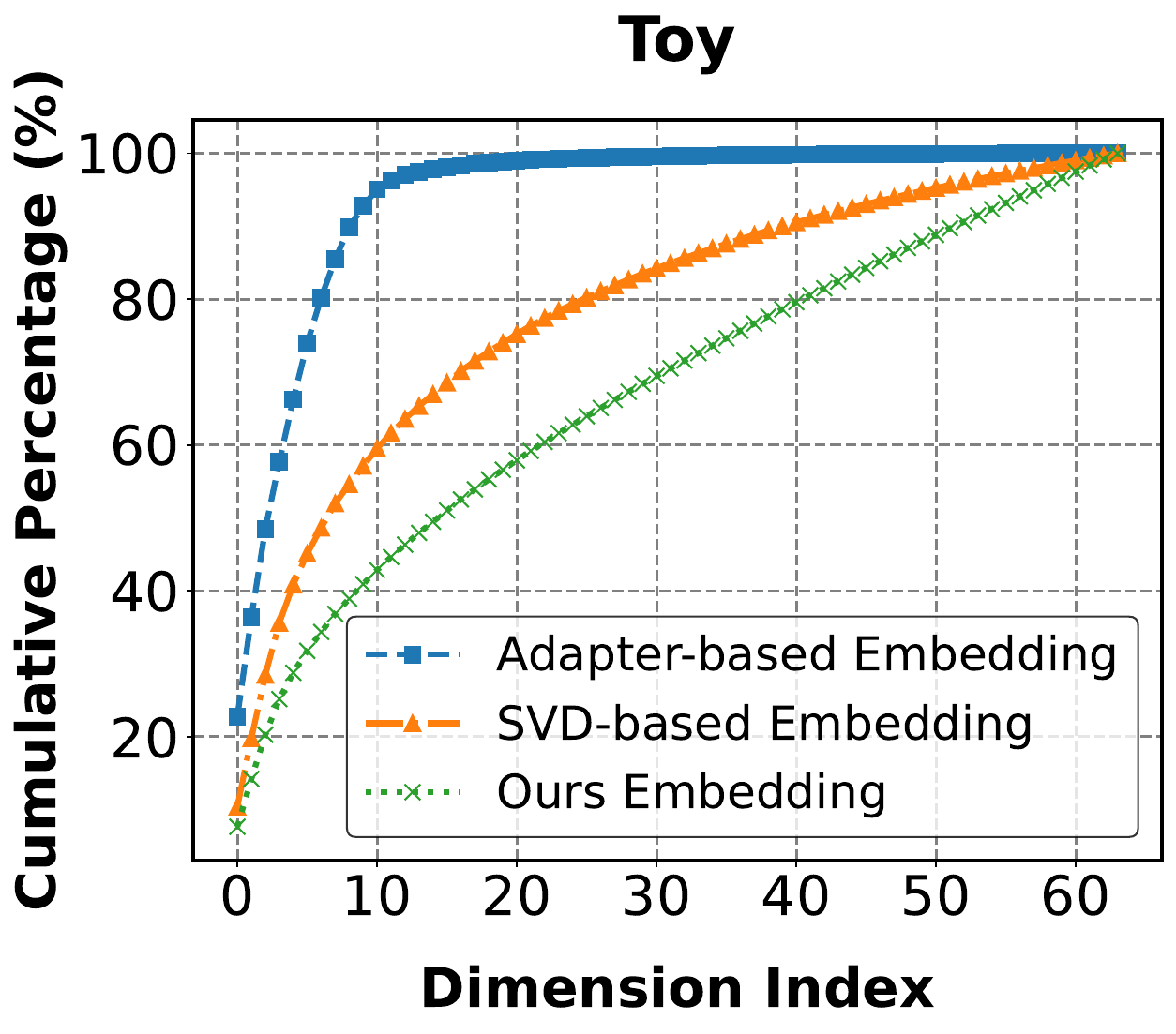}
    \end{subfigure}        
    \begin{subfigure}{0.235\textwidth}
        \centering
        \includegraphics[width=\textwidth]{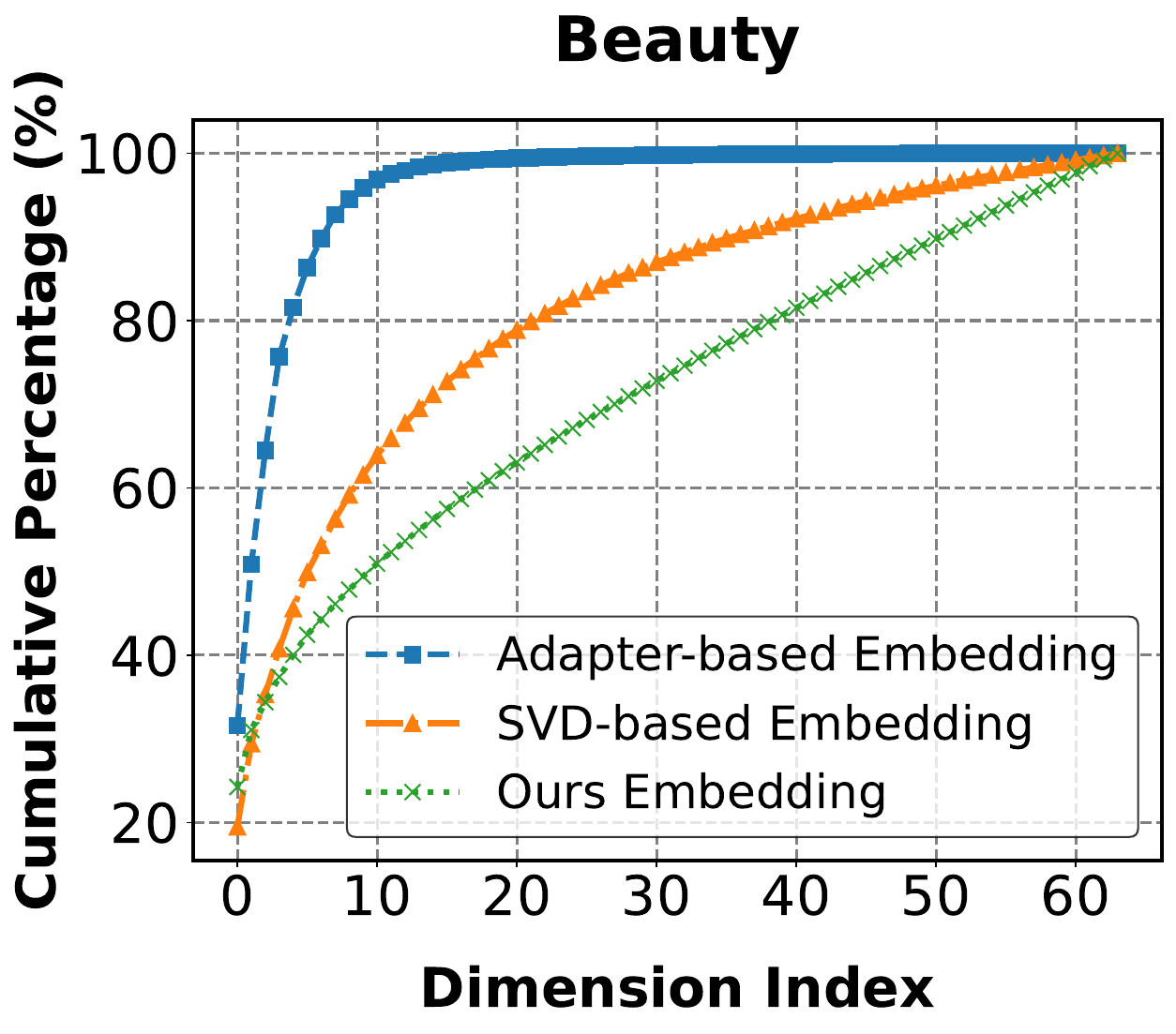}
    \end{subfigure}    
    \vspace{-0.3cm}
    \caption{The cumulative singular values of transformed language embedding matrix on the Toy and Beauty datasets. Similar results are observed on other datasets.} 
    \label{fig:pre_mlp} 
    \vspace{-0.3cm}
\end{figure}

Empirically, we find that SVD‑based methods often outperform projection‑based approaches --- \eg AlphaFuse~\cite{hu2025alphafuse} frequently surpass recent adapters as reported in Table~\ref{tab:main}.  This is surprising given that SVD-based methods are static and hand-crafted. Notably, SVD-based methods typically utilize only a few principal spectral components, discarding rich information contained in the remaining spectrum. By contrast, adapter-based methods are more flexible and adaptive, aligning with the recommendation objective and offering higher potential. To understand this counterintuitive result, we analyze existing adapters and observe pronounced spectral dimension collapse  (\cf Figure~\ref{fig:pre_mlp}):  most eigenvalues of the projected embeddings approach zero, yielding a low-rank representation that carries limited semantic signal concentrated in a few dominant directions. Similar collapse phenomena have been reported for MLP architectures in other domains~\cite{jing2021understanding,chen2024towards}, but it is particularly pronounced in SR adapters. This naturally raises a key question: \textbf{\textit{Can we design a projector that preserves the adaptability of learned adapters while explicitly mitigating spectral collapse?}} 

Towards this end, we propose \textbf{SpecTran}, a novel \underline{\textbf{Spec}}tral-aware \underline{\textbf{Tran}}sformer-based  adapter that adaptively projects language embeddings into the item space.  SpecTran draws on the strengths of SVD‑based strategies by operating in the spectral domain, but overcomes their rigidity via a learnable module that selects and aggregates informative spectral components under the SR objective. Unlike static spectral pruning such as SVD-based strategies, SpecTran attends over the full spectrum, and enables each output dimension to be fully utilized for incorporating informative spectral components. Furthermore, because a vanilla transformer is agnostic to spectral salience in the semantic space, we introduce a learnable spectral positional encoding that injects singular‑value information as an inductive bias, guiding attention toward principal spectral components. Moreover, different output dimensions are encouraged to exploit diverse principal spectral components.

Beyond effectiveness, SpecTran offers practical advantages: it is lightweight, introducing only a modest number of additional trainable parameters and incurring limited extra computational overhead; and it is model-agnostic, enabling integration with a wide range of SR backbones. We evaluate SpecTran on four real-world datasets across three representative SR backbones and observe consistent improvements over state-of-the-art baselines, with an average improvement of 9.17\%.

\begin{itemize}[leftmargin=*]
\item We identify and analyze key limitations of recent language-to-item embedding transformation strategies, highlighting spectral collapse in adapter-based methods.

\item We propose SpecTran, a spectral-aware adapter that leverages a transformer to adaptively select and fuse important spectral information, mitigating dimension collapse and enabling more comprehensive spectral exploitation.

\item We conduct extensive experiments to validate the effectiveness of SpecTran over state-of-the-art methods with minimal extra training parameters and computational overhead.

\end{itemize}

\section{Preliminary}
\label{sec:Preliminary}
\subsection{Task Formulation} 
This work mainly focus on sequential recommendation~\cite{sun2019bert4rec,kang2018self,yang2023generic,boka2024survey}, which has secured a pivotal role in various modern recommendation systems. Sequential recommendation aims to deduce users' dynamic preferences based on their historical  interactions. Given a user set $\mathcal{U}$ with $M$ users and an item set $\mathcal{I}$ with $N$ items,  a user's historical interactions can be organized in chronological order $ S_u = (i_1, i_2, \ldots, i_{t-1})$ where $i_{k} \in \mathcal{I}$ denotes the $k$-th item that the user interacted with. 
The task of sequential recommendation is to predict the next item $i_{t}$ that the user is likely to interact with based on the historical interactions.


Embedding-based models are conventionally employed for SR~\cite{kang2018self,hidasi2015session,xie2022contrastive,yang2023generic}. These methods often assign each item as an independent numerical ID, which is then mapped to a learnable ID embedding table expressed as $\mathbf{E}_{\text{id}} = \{\mathbf{e}_i\}_{i=1}^N \in \mathbb{R}^{N \times d}$, where $d$ denotes the dimension of the ID embeddings (\eg 64 or 128). Various sequential recommendation architectures, \eg LSTMs~\cite{yu2019review,hidasi2015session} and Transformers~\cite{khan2022transformers,sun2019bert4rec}, have been used to learn item ID embeddings and to model interactions among items.
Although effective, these traditional sequential recommendation methods rely solely on item ID information while ignoring the rich textual information of items (e.g., titles, descriptions). In fact, this information explicitly conveying item properties, providing considerable potential to enhance recommendation accuracy.


\subsection{Language Embedding Enhancement} 
Benefiting from the abundant open-world knowledge and strong semantic understanding capabilities of LLMs~\cite{achiam2023gpt,dubey2024llama}, existing studies have explored leveraging LLMs to encode item textual information and inject it into traditional sequential recommendation methods~\cite{xi2024towards,liu2024llm,sun2024large,ren2024enhancing,ren2024representation}. These methods can be summarized in the following three steps:

\subsubsection{Semantic Encoding}  
The textual descriptions of items contain rich semantic information, which can be encoded into semantic embeddings by the LLM. By encoding these natural language texts with an LLM, we can obtain high-dimensional LLM semantic embeddings for the items. Let $l$ denote the embedding dimension of the LLM. The item semantic embeddings can be denoted as $\mathbf{E}_{\text{LLM}}  \in \mathbb{R}^{N \times l}$. The LLM encoding process can be formulated as:
\begin{equation}
\mathbf{E}_{\text{LLM}} = \text{LLMEnc}(\mathcal{T}),
\end{equation}
where  $\mathcal{T}$ denotes the natural language descriptions of all items, and $\text{LLMEnc}(\cdot)$ refers to the LLM used for textual encoding (\eg text-embedding-3~\footnote{https://platform.openai.com/docs/guides/embeddings}, LLaMA3-8B~\cite{dubey2024llama}).


\subsubsection{Embedding Transformation}
Notably, Language embeddings are high-dimensional (e.g., 4096) representations of semantic content, whereas item embeddings are typically lower-dimensional capturing collaborative signals. To tackle such space gap, various embedding transformation strategies have been developed. 
They target at distilling useful semantic from language models to enrich the item ID embeddings. Existing approaches can be broadly divided into two categories:

\textbf{(1) Adapter‑based methods.} The most common approach is to use a learnable MLP (Multi-Layer Perceptron) as an adapter for embedding projection~\cite{yuan2023go,hou2022towards,ren2024representation,liu2024llm}. The process can be formally expressed as:  
\begin{equation}
\textbf{E}_{s} = f_{\theta}^\text{MLP}(\textbf{E}_{\text{LLM}}),
\end{equation} 
where $\textbf{E}_{s} \in \mathbb{R}^{N\times d}$ is the semantic embedding after transformation, $f_{\theta}^\text{MLP}(\cdot)$ is the MLP with learnable parameters $\theta$.  

The advantage of this method lies in its ability to optimize the learnable network using the recommendation objective loss, thereby adaptively extracting information from the original semantic embeddings towards better recommendation performance. However, in practice, we find such MLP-based adapters often do not yield satisfactory performance. Typically, as reported in the Table~\ref{tab:main}, we find most Adapter-based methods perform worse than the state-of-the-art SVD-based methods, and sometimes their performance is even weaker than the backbone model (\eg the Toy dataset of the HSTU backbone). These results are quite impressive, given that adapter is adaptive and flexible, aligning with prevailing design principles in modern AI.


We conduct an in-depth analysis of the item embeddings after Adapter‑based transformation to understand the underlying reason for their limited effectiveness. The result is illustrated in Figure~\ref{fig:pre_mlp}. We observe that the singular values corresponding to the top 10 dimensions of the covariance matrix of the reduced item language embeddings $\textbf{E}_{s}$ account for approximately 95\% of the total singular value. This reveals a severe \textbf{spectral dimension collapse}~\cite{chen2024towards} phenomenon \ie, information is concentrated in only a few dimensions, while most dimensions are useless for the recommendation model.

\begin{figure*}[t]
    \centering 
    \includegraphics[width=0.975\textwidth]{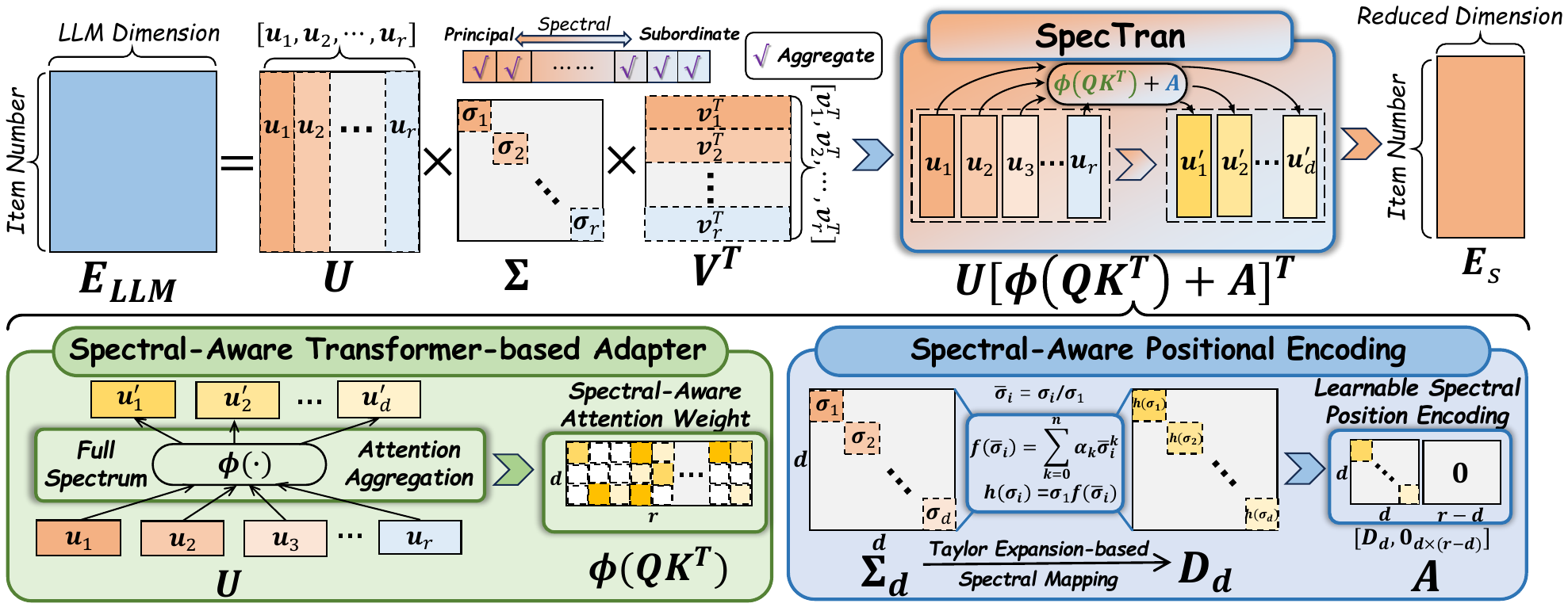} 
    \caption{The overall framework of the proposed SpecTran.} 
    \label{fig:framework} 
\end{figure*}

\textbf{(2) SVD-based method.} Another perspective is to perform transformation via Singular Value Decomposition (SVD)~\cite{hoecker1996svd}.  This approach decomposes the item semantic embeddings into the spectral domain and keeps the top-$d$ largest singular values for transformation~\cite{zhang2024id,zhang2025llminit,hu2025alphafuse}.  Formally, given the item semantic embeddings $\mathbf{E}_{\text{LLM}} \in \mathbb{R}^{N \times l}$ encoded by the LLM, the SVD process is:  
\begin{equation}
\mathbf{E}_{\text{LLM}} = \mathbf{U} \mathbf{\Sigma} \mathbf{V}^T,
\label{eq:svd}
\end{equation}
where $\mathbf{U} \in  \mathbb{R}^{N \times r}$ denotes the left singular matrix, and each column vector represents a spectral component in the spectral space. (Here $r$ is the rank of $\mathbf{E}_{LLM}$ and is typically equal to the language embedding dimension $l$.)
$\mathbf{\Sigma} = \mathrm{diag}(\sigma_1, \sigma_2, \cdots, \sigma_r) \in \mathbb{R}^{r \times r}$ represents the singular value magnitudes with  $\sigma_1 \ge \sigma_2 \ge \cdots \ge \sigma_r \ge 0$,  and $\mathbf{V} \in \mathbb{R}^{l \times r}$ denotes the right singular matrix. 

The transformation process is accomplished by retaining a small number of principal  spectral components while discarding the subordinate ones.  
Formally, this can be expressed as:  
\begin{equation}
\label{eq:svd_trunc}
\begin{aligned}
\mathbf{E}_{s} 
&= \mathbf{E}_{\text{LLM}} \; \mathbf{V}[:, :d] \\&= \mathbf{U} \mathbf{\Sigma} \mathbf{V}^T \mathbf{V}[:, :d] \\
&= \mathbf{U}
\begin{bmatrix}
\mathbf{\Sigma}_d \\
\mathbf{0}_{(r-d) \times d}
\end{bmatrix},
\end{aligned}
\end{equation}
where $\mathbf{V}[:, :d]$ denotes selecting the top-$d$ principal  spectral components with the large singular values,   
$\mathbf{0}_{(r-d) \times d}$ is the zero matrix, and  
$\mathbf{\Sigma}_d = \mathrm{diag}(\sigma_1, \sigma_2, \cdots, \sigma_d)$ represents the diagonal matrix composed of the largest $d$ singular values.  

Although SVD-based methods have shown certain effectiveness, they still suffer from two critical weaknesses:


\textbf{i) Information loss in subordinate spectral space.} 
Although the subordinate spectral components contain "less semantic information", this  conclusion is drawn from the LLM’s semantic space perspective.  
From the perspective of the collaborative space in traditional recommendation models, subordinate spectral components may still convey useful semantic cues.  
Therefore, simply discarding this large portion of subordinate spectral components may reduce the efficiency of semantic information utilization.  
To address this issue, we propose a global spectral attention mechanism for subordinate spectral component aggregation with sparsified activation function, which aims to effectively aggregate the overlooked subordinate spectral components thereby improving the quality of semantic knowledge.

\textbf{ii) Static and hand-crafted spectral weights.}
Previous studies allocate spectral weights with static approaches such as the variance characteristics~\cite{zhang2025llminit} or singular values~\cite{zhang2024id}.
Typically, the most advanced AlphaFuse~\cite{hu2025alphafuse} normalizes the singular values of the principal  spectral components to the identity matrix for improvement:  
\begin{equation}
\begin{aligned}
\mathbf{E}_{s} 
&= \mathbf{U} \mathbf{\Sigma} \mathbf{V}^T  \mathbf{V}[:, :d] \mathbf{\Sigma}^{-1}[:, :d] \\
&= \mathbf{U}
\begin{bmatrix}
\mathbf{I}_d \\
\mathbf{0}_{(r-d) \times d}
\end{bmatrix},
\end{aligned}
\label{eq:svd_hand}
\end{equation}   
where $\mathbf{I}_d$ is the $d \times d$ identity matrix. However, these static or hand-crafted methods lack flexibility and fail to effectively distinguish the importance of each spectral dimension, resulting in limited recommendation performance.
Therefore, we propose a learnable Transformer-based spectral aggregation strategy, allowing a more flexible manner for learning spectral weights.

\subsubsection{Embedding Fusion}  
After obtaining the semantic embedding from transformation, it is necessary to fuse it with the original ID embedding, injecting the semantic information into the recommendation model. The fusion process can be expressed  as:  
\begin{equation}
\mathbf{E}_{\text{item}} = \text{Fusion}(\mathbf{E}_{s}, \mathbf{E}_{id}),
\end{equation}
where $\mathbf{E}_{\text{item}} \in \mathbb{R}^{N \times d}$ denotes the final item representation, which integrates both the ID embeddings and the semantic embeddings. $\text{Fusion}(\cdot)$ refers to various fusion strategies, such as concatenation~\cite{liu2024llm}, element-wise addition~\cite{hou2022towards}, semantic initialization~\cite{zhang2025llminit} and semantic null-space learning~\cite{hu2025alphafuse}. In our experiments, we follow their original paper setting or adopt the optimal fusion strategy for each compared method to ensure their best performance.  

\section{Methodology}
In this section, we introduce the proposed \textbf{SpecTran}, a novel Spectral-aware Transformer-based adapter that adaptively projects language embeddings into the item space.   
SpecTran operates directly in the spectral domain, replacing the static spectral filtering in traditional SVD-based methods with a learnable attention mechanism. This enables the adaptive aggregation of informative spectral components under the recommendation objective.
The overall framework is illustrated in Figure~\ref{fig:framework}.

\subsection{Spectral-Aware Transformer-based Adapter}
SpecTran draws on the strengths of SVD‑based strategies by operating in the spectral domain, while mitigating their rigidity with a learnable module that selects and aggregates informative spectral components under the recommendation objective. Specifically, given the SVD decomposition of item language embeddings in \Cref{eq:svd}, we have the spectral space $\mathbf{U} = [\mathbf{u}_1,\mathbf{u}_2,\cdots,\mathbf{u}_r]\in \mathbb{R}^{N \times r}$, which contains all $r$ (\eg 4096) spectral components. Instead of statically selecting only the top $d$ (\eg 64) principal spectral components, SpecTran performs a learnable aggregation over the full spectrum.

\textbf{Spectral-Aware Attention.} Inspired by the attention mechanism's ability to adaptively aggregate informative information, we develop a specialized Transformer-based adapter.  Intuitively, each output dimension of the adapter attends over the entire spectral space, leveraging the attention mechanism to select the most informative components. Specifically, the adapter is formulated as follows:
\begin{equation}
\mathbf{E}_{s} = \mathbf{U}[\phi({\mathbf{QK}^T})]^T,
\label{eq:spec_basic}
\end{equation}
where $\mathbf{Q} \in \mathbb{R}^{d \times r}$ denotes the learnable Query matrix that captures the characteristics of each transformed output dimension; $\mathbf{K}\in \mathbb{R}^{r \times r}$ denotes the learnable Key matrix that captures the features of each spectral dimension; $\mathbf{U} \in \mathbb{R}^{N \times r}$ acts as the Value matrix directly derived from the spectral space; and $\phi(\cdot)$ denotes the activation function. The term $\phi(\mathbf{Q}\mathbf{K}^T)$ calculates the attention scores of each output dimension over the full spectrum, which is then transposed to aggregate the Value matrix $\mathbf U$.

Notably, we do not adopt the vanilla Transformer architecture, which computes Query, Key, and Value matrices through the input-dependent projections on the spectral matrix $U$  (\ie $\mathbf{W}_Q\mathbf{U}$, $\mathbf{W}_K\mathbf{U}$ and $ \mathbf{W}_V\mathbf{U}$, where $\mathbf{W}_Q,\mathbf{W}_K,\mathbf{W}_V\in\mathbb{R}^{d \times N}$). This design choice is motivated by two reasons: First, in large-scale recommendation scenarios, the number of items $N$ is typically massive (\eg millions). Introducing large-scale matrix multiplications along the $N$ dimension would drastically increase parameter complexity and incur unaffordable inference latency. Second, in our scenario, each query corresponds to a specific output dimension rather than an input token. Although its implementation differs from a vanilla Transformer, the core mechanism remains consistent: it learns adaptive queries and keys to compute attention scores, thereby aggregating relevant spectral information for each output dimension.

\textbf{Sparsified Activation.} For SpecTran, we observe that the vanilla Softmax activation is sub-optimal (\cf Table~\ref{tab:ab}). The learned attention weights tend to be relatively flat and uniformly distributed, which renders the adapter ineffective as a spectral selector. Typically, given the high dimensionality of the original spectrum, the cumulative attention weights over the relatively less important dimensions (\eg components from $d+1$ to $r$) can easily dominate the signals from the top $d$ principal components, causing the spectrum selection to fail. To alleviate this issue, we adopt Softshrink~\cite{donoho1995noising} as a sparsified activation function: 
\begin{equation}
\text{Softshrink}(x) =
\begin{cases}
x - \lambda, & x > \lambda, \\
x + \lambda, & x < -\lambda, \\
0, & \text{otherwise},
\end{cases}
\end{equation}
where $\lambda$ is the shrinkage threshold. Unlike standard Softmax attention, we do not enforce attention normalization. Instead, Softshrink acts as a sparse spectral gating operator, allowing flexible magnitude modeling beyond probability constraints. It explicitly suppresses small-magnitude components and encourages sparsity in attention weights, effectively serving as a sparse spectral filter. This design prevents subordinate spectral components with weak signals from overwhelming dominant principal components (\cf Table~\ref{tab:case}) and stabilizes optimization.

Comparing \Cref{eq:spec_basic} with the truncated SVD form in \Cref{eq:svd_trunc,eq:svd_hand}, we observe that SpecTran replaces the fixed diagonal spectral filter matrix with a learnable spectral-aware Transformer-based adapter $[\phi({\mathbf{QK}^T})]^T$. When the Transformer attention result degenerates to 
$
\begin{bmatrix}
\mathbf{\Sigma}_d \\
\mathbf{0}
\end{bmatrix}
$
or 
$
\begin{bmatrix}
\mathbf{I}_d \\
\mathbf{0}
\end{bmatrix},
$
SpecTran reduces to existing SVD-based methods. Therefore, our formulation naturally generalizes previous spectral truncation strategies.

\subsection{Spectral-Aware Positional Encoding}
Although the proposed adapter adaptively aggregates information from the entire spectral space, it cannot intrinsically perceive the inherent importance of different spectral components from the original language space. To incorporate this prior knowledge, we introduce the Spectral-Aware Positional Encoding module. We aim to utilize the singular values of the principal spectral components as an inductive bias, guiding the model to effectively learn from the language space.

Our design is guided by three principles: 1) Given the potentially important role of principal components, we inject bias terms exclusively into the top $d$ spectral dimensions. This highlights their importance while reducing learning complexity. 2) To encourage the output dimensions to capture diverse information, the positional encodings for different output dimensions should be distinct, and preferably orthogonal. 3) The positional encodings require flexibility. Existing SVD-based methods rigidly use singular values as static weights, which may not directly quantify their true importance for recommendation tasks. To address this, we construct a flexible learnable mapping function to transform singular values into task-specific importance weights.

\textbf{Learnable Spectral Positional Encoding.} Guided by these principles, we formulate the spectral positional encoding matrix $\mathbf{A} \in \mathbb{R}^{d \times r}$  as:
\begin{equation}
\mathbf{A}=[\mathbf{D}_d, \mathbf{0}_{d \times (r-d)}],
\end{equation}
where  $\mathbf{D}_d \in \mathbb{R}^{d \times d}$ denotes the bias terms introduced on the $d$-dimensional principal spectral components.  Notably, $\mathbf{D}_d$ is a diagonal matrix where each diagonal element $h(\sigma_i)$ is a transformed value derived from the original singular values via a learnable function:
\begin{equation}
\mathbf{D}_d=\mathrm{diag}\{h(\sigma_1), \dots, h(\sigma_d)\}.
\end{equation}
Intuitively, this positional encoding encourages each output dimension to prioritize the information from corresponding principal spectral components, with the magnitude of this preference modulated by their original singular values.

\textbf{Taylor Expansion-based Spectral Mapping.}
To flexibly model the relationship between singular values and task-specific importance, we design $h(\cdot)$ as a highly expressive function. Leveraging the universal approximation capability of the Taylor expansion, we express $h(\cdot)$ as a Taylor polynomial of the original spectral singular values with learnable scaling coefficients.  To ensure numerical stability and facilitate coefficient learning, we introduce a normalized singular value $\bar \sigma_i$ and learn the polynomial over this normalized space:
\begin{equation} 
\begin{aligned} 
\bar{\sigma}_i &= \sigma_i / \sigma_1, 
\\ f(\bar{\sigma}_i) &= \alpha_0 + \alpha_1 \bar{\sigma}_i + \alpha_2 \bar{\sigma}_i^2 + \cdots + \alpha_n \bar{\sigma}_i^n, 
\\ h(\sigma_i) &= \sigma_1 f(\bar{\sigma}_i). 
\end{aligned} 
\end{equation}
We finally multiply the polynomial output by $\sigma_1$ to recover the relative magnitudes.
Overall, the final spectral-aware Transformer-based adapter with learnable position encoding can be expressed as:
\begin{equation}
\mathbf{E}_{s} = \mathbf{U}[\phi({\mathbf{QK}^T})+\mathbf{A}]^T.
\label{eq:spectran}
\end{equation} 
With this design, SpecTran can adaptively select and aggregate important spectral information, mitigating dimension collapse and enabling more
comprehensive spectral exploitation.

\section{Experiments}
We aim to answer the following research questions:
\begin{itemize}[leftmargin=*]
  \item $\mathbf{RQ1:}$ How does SpecTran perform compared with existing state-of-the-art methods?
  \item $\mathbf{RQ2:}$ What are the impacts of different components of SpecTran?
  \item $\mathbf{RQ3:}$ How do different hyperparameters affect SpecTran?
  \item $\mathbf{RQ4:}$ What is the training and inference efficiency of SpecTran?
  \item $\mathbf{RQ5:}$ What characteristics do the learned principal  and subordinate spectral weights exhibit in SpecTran?
\end{itemize}

\subsection{Experimental Settings}
\subsubsection{Datasets}
Four well-known real-world datasets: \textit{Amazon Toys and Games}, \textit{Amazon Beauty}, \textit{Amazon Clothing, Shoes and Jewelry} and \textit{Amazon Office Products} are utilized in our experiments~\footnote{\url{https://cseweb.ucsd.edu/~jmcauley/datasets/amazon/links.html}}, which are commonly used for the studies of LLM-enhanced recommendation~\cite{xi2024towards,zhang2025llminit,cui2024distillation}. 
For fair comparisons, we closely adhered to the preprocessing methods used in recent work~\cite{hu2025alphafuse}: 
\textbf{Leave-One-Out} setting is applied to segment the user interaction sequences (\ie the last item in a user’s interaction sequence is designated as the next item for prediction and the remaining items are used as historical interactions for training).  The user sequences are then sorted in ascending order by timestamp and split into training, validation, and testing sets with an 8:1:1 ratio. The dataset statistics are presented in Table~\ref{tab:datasets}.

\begin{table}[]
\centering
\caption{Statistics of the datasets.}
\vspace{-0.3cm}
\label{tab:datasets}
\begin{tabular}{@{}c|cccc@{}}
\toprule
\textbf{Dataset} & \textbf{Toy} & \textbf{Beauty} & \textbf{Clothing} & \textbf{Office} \\ \midrule
\textbf{\#User} & 19,124 & 22,332 & 39,230 & 4,895 \\
\textbf{\#Item} & 11,757 & 12,086 & 22,948 & 2,414 \\
\textbf{\#Interaction} & 141,630 & 168,446 & 266,481 & 41,462 \\
\textbf{Density} & 0.0630\% & 0.0624\% & 0.0296\% & 0.3509\% \\ \bottomrule
\end{tabular}
\vspace{-0.3cm}
\end{table}

\subsubsection{Backbones}
To validate the effectiveness and generalization capability of SpecTran, we conduct evaluations on three well-known representative traditional sequential recommendation methods: 

\begin{itemize}[leftmargin=*]
    \item \textbf{BERT4Rec (CIKM'19)~\cite{sun2019bert4rec}} is a representative sequential recommender inspired by BERT~\cite{devlin2019bert}. It applies bidirectional self-attention to capture both past and future interactions within user sequences with a Cloze-style training objective.
    \item \textbf{SASRec (ICDM'18)~\cite{kang2018self}} is a well-known Transformer-based sequential recommender. It leverages causal self-attention mechanisms to capture collaborative signals among historical items and predict the next item based on users' interaction sequences.
    \item \textbf{HSTU (ICML'24)~\cite{zhai2024actions}} is a novel generative recommender for large-scale recommendation. It adopts pointwise aggregated attention instead of the softmax-based attention, and incorporates a temporal-aware relative bias into the attention scores.
\end{itemize}
Following the previous work~\cite{hu2025alphafuse}, we uniformly adopt the InfoNCE loss~\cite{oord2018representation} with 64 negative samples as the recommendation loss for all backbones to ensure fairness in evaluation.
 
\subsubsection{Baselines}
We compare SpecTran with the following semantic embedding transformation methods, including:

\begin{itemize}[leftmargin=*]
  \item \textbf{Adapter‑based methods:} \textbf{MoRec (SIGIR'23)}~\cite{yuan2023go} adopts pre-trained language models to encode item representations, followed by a dense MLP layer for the dimension transformation. \textbf{UniSRec (KDD'22)}~\cite{hou2022towards} utilizes a mixture-of-experts (MoE) adapter to refine language-based embeddings, generating the final item vectors with improved adaptability. \textbf{RLMRec (WWW'24)}~\cite{ren2024representation} aligns the dimensions of item semantic embeddings and ID embeddings through an MLP, with a reconstruction loss introduced to enhance the alignment learning process. \textbf{LLM-ESR (NIPS'24)}~\cite{liu2024llm} introduces a dual-view framework combining semantic initialization with adaptive projection, aiming to alleviate the long-tailed user/item problem.
  \item \textbf{SVD-based methods: } \textbf{WhitenRec (ICDE'24)}~\cite{zhang2024id} applies a spectral-based whitening transformation on pre-trained item textual embeddings, which are subsequently fed into an adapter to yield the final representations. \textbf{LLMInit (arXiv'25)}~\cite{zhang2025llminit} selects the top-$k$ dimensions with the highest variance of item semantic embeddings to initialize ID embeddings. \textbf{AlphaFuse (SIGIR'25)}~\cite{hu2025alphafuse} presents a state‑of‑the‑art SVD-based transformation approach by learning item ID embeddings within the null space of semantic embeddings, effectively mitigating the semantic gap.
\end{itemize}

\subsubsection{Evaluation Metrics}
We employ two widely-used metrics $HR@K$ and $NDCG@K$ to evaluate the performance ($K=10,20$) following previous work~\cite{wang2025msl}. HR (Hit Ratio) evaluates whether a relevant item appears in the top-K recommendations. NDCG (Normalized Discounted Cumulative Gain) evaluates recommendation quality accounting for both relevance scores and their positions.

\subsubsection{Implementation Details}
Following previous work~\cite{yang2023generic}, we implement traditional recommendation backbones by Adam \cite{kingma2014adam} optimizer with a learning rate of 0.001, an embedding dimension of 128 and a batch size of 256. We tune the weight decay in \{1e-4, 1e-5, 1e-6, 0\} and the dropout ratio among $[0, 0.5]$ in the step of 0.1 to avoid overfitting. We also set the maximum user sequence length to 10 following~\cite{yang2023generic, hu2025alphafuse}.
For \textbf{BERT4Rec}, we follow the original cloze-style training paradigm and search the masking ratio within [0.2, 0.4, 0.6]. For \textbf{SASRec}, we adopt 2 Transformer blocks with 1 attention head, as suggested in the original paper. For \textbf{HSTU}, we only use the ID features of items to fit the setting of sequential recommendation.
For all the baselines, we closely followed the settings suggested by their original papers and finely tuned their hyperparameters to ensure their optimal performance. Unless otherwise specified, we use LLaMA3-8B~\cite{dubey2024llama} as the LLM to encode item titles for semantic embeddings.

For our method, we initialize the Taylor expansion order coefficients $\{\alpha_k\}_{k=0}^{n}$ as 1 to keep stable training. 
For the Query and Key matrices $\textbf{Q}$ and $\textbf{K}$, we initialize them with a Gaussian distribution $\mathcal{N}(0, 0.1^2)$ to ensure sparsity. We assign the sparsified threshold $\lambda$ of the Softshrink activation function as a learnable parameter and initialized as 0.
We set the max training epoch number at 200 and evaluate the performance per epoch. Early stop strategy~\cite{prechelt2002early} is used on the $NDCG@20$ with patience as 10. All methods are implemented in PyTorch and run on 8 Nvidia 4090 GPUs. Our code is available at here~\footnote{\url{https://github.com/istarryn/SpecTran}}.

\begin{table*}[]
\centering
\caption{The performance comparison on four real-world datasets. The best and the second performances are bolded and underlined, respectively. "Impr." denotes the improvement of SpecTran over the best baseline method. "N" represents NDCG, and "H" represents Hit Ratio.}
\vspace{-0.3cm}
\label{tab:main}
\scalebox{0.75}{
\begin{tabular}{@{}c|c|cccc|cccc|cccc|cccc@{}}
\toprule
 &  & \multicolumn{4}{c|}{\textbf{Toy}} & \multicolumn{4}{c|}{\textbf{Beauty}} & \multicolumn{4}{c|}{\textbf{Clothing}} & \multicolumn{4}{c}{\textbf{Office}} \\ \cmidrule(l){3-18} 
\multirow{-2}{*}{\textbf{Backbone}} & \multirow{-2}{*}{\textbf{Method}} & N@10 & N@20 & H@10 & H@20 & N@10 & N@20 & H@10 & H@20 & N@10 & N@20 & H@10 & H@20 & N@10 & N@20 & H@10 & H@20 \\ \midrule
 & Base & 0.0189 & 0.0247 & 0.0381 & 0.0614 & 0.0214 & 0.0286 & 0.0429 & 0.0716 & 0.0071 & 0.0098 & 0.0149 & 0.0256 & 0.0413 & 0.0549 & 0.0805 & 0.1346 \\
 & MoRec & 0.0252 & 0.0324 & 0.0499 & 0.0784 & 0.0238 & 0.0325 & 0.0479 & 0.0826 & 0.0114 & 0.0149 & 0.0221 & 0.0360 & 0.0414 & 0.0561 & 0.0840 & 0.1422 \\
 & UniSRec & 0.0235 & 0.0311 & 0.0478 & 0.0780 & 0.0245 & 0.0328 & 0.0493 & 0.0821 & 0.0111 & 0.0148 & 0.0224 & 0.0369 & 0.0433 & 0.0563 & 0.0841 & 0.1425 \\
 & LLM-ESR & 0.0271 & 0.0356 & 0.0558 & 0.0898 & 0.0242 & 0.0320 & 0.0474 & 0.0786 & 0.0129 & 0.0167 & 0.0253 & 0.0405 & 0.0337 & 0.0471 & 0.0705 & 0.1240 \\
 & RLMRec & 0.0239 & 0.0314 & 0.0475 & 0.0773 & 0.0266 & 0.0347 & 0.0548 & 0.0868 & 0.0119 & 0.0155 & 0.0232 & 0.0378 & {\ul 0.0436} & {\ul 0.0567} & {\ul 0.0883} & {\ul 0.1426} \\

  & WhitenRec & 0.0237 & 0.0308 & 0.0500 & 0.0782 & 0.0239 & 0.0329 & 0.0497 & 0.0854 & 0.0103 & 0.0135 & 0.0208 & 0.0335 & 0.0422 & 0.0551 & 0.0838 & 0.1348 \\
 & LLMInit & 0.0257 & 0.0338 & 0.0541 & 0.0858 & 0.0236 & 0.0313 & 0.0461 & 0.0768 & 0.0103 & 0.0140 & 0.0212 & 0.0359 & 0.0412 & 0.0553 & 0.0844 & 0.1410 \\
 & AlphaFuse & {\ul 0.0274} & {\ul 0.0367} & {\ul 0.0583} & {\ul 0.0952} & {\ul 0.0268} & {\ul 0.0365} & {\ul 0.0565} & {\ul 0.0952} & {\ul 0.0131} & {\ul 0.0182} & {\ul 0.0275} & {\ul 0.0478} & 0.0421 & 0.0554 & 0.0854 & 0.1385 \\
 & \cellcolor[HTML]{96FFFB}SpecTran & \cellcolor[HTML]{96FFFB}\textbf{0.0321} & \cellcolor[HTML]{96FFFB}\textbf{0.0415} & \cellcolor[HTML]{96FFFB}\textbf{0.0674} & \cellcolor[HTML]{96FFFB}\textbf{0.1049} & \cellcolor[HTML]{96FFFB}\textbf{0.0281} & \cellcolor[HTML]{96FFFB}\textbf{0.0381} & \cellcolor[HTML]{96FFFB}\textbf{0.0596} & \cellcolor[HTML]{96FFFB}\textbf{0.0994} & \cellcolor[HTML]{96FFFB}\textbf{0.0139} & \cellcolor[HTML]{96FFFB}\textbf{0.0193} & \cellcolor[HTML]{96FFFB}\textbf{0.0299} & \cellcolor[HTML]{96FFFB}\textbf{0.0516} & \cellcolor[HTML]{96FFFB}\textbf{0.0489} & \cellcolor[HTML]{96FFFB}\textbf{0.0629} & \cellcolor[HTML]{96FFFB}\textbf{0.0983} & \cellcolor[HTML]{96FFFB}\textbf{0.1540} \\ \cmidrule(l){2-18} 
\multirow{-10}{*}{BERT4Rec} & Impr. & {\color[HTML]{C00000} \textbf{17.15\%}} & {\color[HTML]{C00000} \textbf{13.08\%}} & {\color[HTML]{C00000} \textbf{15.61\%}} & {\color[HTML]{C00000} \textbf{10.19\%}} & {\color[HTML]{C00000} \textbf{4.85\%}} & {\color[HTML]{C00000} \textbf{4.38\%}} & {\color[HTML]{C00000} \textbf{5.49\%}} & {\color[HTML]{C00000} \textbf{4.41\%}} & {\color[HTML]{C00000} \textbf{6.11\%}} & {\color[HTML]{C00000} \textbf{6.04\%}} & {\color[HTML]{C00000} \textbf{8.73\%}} & {\color[HTML]{C00000} \textbf{7.95\%}} & {\color[HTML]{C00000} \textbf{12.16\%}} & {\color[HTML]{C00000} \textbf{10.93\%}} & {\color[HTML]{C00000} \textbf{11.33\%}} & {\color[HTML]{C00000} \textbf{7.99\%}} \\ \midrule
 & Base & 0.0246 & 0.0281 & 0.0482 & 0.0626 & 0.0239 & 0.0295 & 0.0496 & 0.0719 & 0.0079 & 0.0096 & 0.0161 & 0.0230 & 0.0309 & 0.0423 & 0.0674 & 0.1130 \\
 & MoRec & 0.0269 & 0.0331 & 0.0561 & 0.0805 & 0.0260 & 0.0328 & 0.0520 & 0.0793 & 0.0107 & 0.0146 & 0.0218 & 0.0372 & 0.0348 & 0.0468 & 0.0752 & 0.1230 \\
 & UniSRec & 0.0261 & 0.0332 & 0.0550 & 0.0834 & 0.0268 & 0.0344 & 0.0548 & 0.0849 & 0.0108 & 0.0144 & 0.0222 & 0.0365 & 0.0399 & 0.0522 & 0.0801 & 0.1297 \\
 & LLM-ESR & 0.0230 & 0.0295 & 0.0454 & 0.0714 & 0.0233 & 0.0302 & 0.0451 & 0.0724 & 0.0106 & 0.0146 & 0.0218 & 0.0377 & 0.0308 & 0.0395 & 0.0633 & 0.0983 \\
 & RLMRec & 0.0269 & 0.0341 & 0.0565 & 0.0852 & 0.0267 & 0.0347 & 0.0562 & 0.0880 & 0.0111 & 0.0148 & 0.0221 & 0.0371 & 0.0357 & 0.0494 & 0.0797 & 0.1342 \\

  & WhitenRec & 0.0295 & 0.0377 & 0.0597 & 0.0923 & 0.0290 & 0.0377 & 0.0583 & 0.0925 & 0.0134 & 0.0180 & 0.0273 & 0.0459 & {\ul 0.0438} & {\ul 0.0563} & {\ul 0.0866} & {\ul 0.1367} \\
 & LLMInit & 0.0296 & 0.0367 & 0.0614 & 0.0895 & 0.0279 & 0.0355 & 0.0557 & 0.0857 & 0.0112 & 0.0153 & 0.0227 & 0.0392 & 0.0390 & 0.0510 & 0.0789 & 0.1267 \\
 & AlphaFuse & {\ul 0.0358} & {\ul 0.0444} & {\ul 0.0754} & {\ul 0.1094} & {\ul 0.0336} & {\ul 0.0428} & {\ul 0.0680} & {\ul 0.1043} & {\ul 0.0163} & {\ul 0.0215} & {\ul 0.0333} & {\ul 0.0541} & 0.0401 & 0.0511 & 0.0805 & 0.1244 \\
 & \cellcolor[HTML]{96FFFB}SpecTran & \cellcolor[HTML]{96FFFB}\textbf{0.0390} & \cellcolor[HTML]{96FFFB}\textbf{0.0483} & \cellcolor[HTML]{96FFFB}\textbf{0.0856} & \cellcolor[HTML]{96FFFB}\textbf{0.1224} & \cellcolor[HTML]{96FFFB}\textbf{0.0340} & \cellcolor[HTML]{96FFFB}\textbf{0.0437} & \cellcolor[HTML]{96FFFB}\textbf{0.0746} & \cellcolor[HTML]{96FFFB}\textbf{0.1136} & \cellcolor[HTML]{96FFFB}\textbf{0.0175} & \cellcolor[HTML]{96FFFB}\textbf{0.0228} & \cellcolor[HTML]{96FFFB}\textbf{0.0378} & \cellcolor[HTML]{96FFFB}\textbf{0.0588} & \cellcolor[HTML]{96FFFB}\textbf{0.0471} & \cellcolor[HTML]{96FFFB}\textbf{0.0612} & \cellcolor[HTML]{96FFFB}\textbf{0.1003} & \cellcolor[HTML]{96FFFB}\textbf{0.1567} \\ \cmidrule(l){2-18} 
\multirow{-10}{*}{SASRec} & Impr. & {\color[HTML]{C00000} \textbf{8.94\%}} & {\color[HTML]{C00000} \textbf{8.78\%}} & {\color[HTML]{C00000} \textbf{13.53\%}} & {\color[HTML]{C00000} \textbf{11.88\%}} & {\color[HTML]{C00000} \textbf{1.19\%}} & {\color[HTML]{C00000} \textbf{2.10\%}} & {\color[HTML]{C00000} \textbf{9.71\%}} & {\color[HTML]{C00000} \textbf{8.92\%}} & {\color[HTML]{C00000} \textbf{7.36\%}} & {\color[HTML]{C00000} \textbf{6.05\%}} & {\color[HTML]{C00000} \textbf{13.51\%}} & {\color[HTML]{C00000} \textbf{8.69\%}} & {\color[HTML]{C00000} \textbf{7.53\%}} & {\color[HTML]{C00000} \textbf{8.70\%}} & {\color[HTML]{C00000} \textbf{15.82\%}} & {\color[HTML]{C00000} \textbf{14.63\%}} \\ \midrule
 & Base & 0.0257 & 0.0297 & 0.0503 & 0.0660 & 0.0251 & 0.0308 & 0.0509 & 0.0734 & 0.0077 & 0.0095 & 0.0157 & 0.0228 & 0.0326 & 0.0436 & 0.0654 & 0.1091 \\
 & MoRec & 0.0240 & 0.0298 & 0.0477 & 0.0708 & 0.0252 & 0.0325 & 0.0515 & 0.0803 & 0.0102 & 0.0137 & 0.0211 & 0.0349 & 0.0394 & 0.0544 & 0.0819 & 0.1324 \\
 & UniSRec & 0.0236 & 0.0311 & 0.0487 & 0.0785 & 0.0256 & 0.0332 & 0.0508 & 0.0813 & 0.0106 & 0.0143 & 0.0217 & 0.0363 & 0.0396 & 0.0545 & 0.0782 & 0.1383 \\
 & LLM-ESR & 0.0224 & 0.0293 & 0.0440 & 0.0714 & 0.0242 & 0.0312 & 0.0480 & 0.0758 & 0.0101 & 0.0135 & 0.0205 & 0.0340 & 0.0356 & 0.0465 & 0.0691 & 0.1132 \\
 & RLMRec & 0.0233 & 0.0308 & 0.0469 & 0.0767 & 0.0261 & 0.0329 & 0.0510 & 0.0783 & 0.0104 & 0.0139 & 0.0200 & 0.0340 & {\ul 0.0401} & {\ul 0.0550} & {\ul 0.0821} & {\ul 0.1401} \\

  & WhitenRec & {\ul 0.0334} & {\ul 0.0423} & {\ul 0.0702} & {\ul 0.1056} & 0.0251 & 0.0322 & 0.0505 & 0.0789 & 0.0156 & 0.0207 & 0.0319 & 0.0525 & 0.0396 & 0.0519 & 0.0803 & 0.1291 \\
 & LLMInit & 0.0297 & 0.0358 & 0.0608 & 0.0847 & 0.0267 & 0.0343 & 0.0552 & 0.0854 & 0.0114 & 0.0147 & 0.0241 & 0.0374 & 0.0341 & 0.0457 & 0.0754 & 0.1211 \\
 & AlphaFuse & 0.0327 & 0.0415 & 0.0699 & 0.1051 & {\ul 0.0306} & {\ul 0.0397} & {\ul 0.0661} & {\ul 0.1022} & {\ul 0.0157} & {\ul 0.0208} & {\ul 0.0333} & {\ul 0.0533} & 0.0384 & 0.0510 & 0.0776 & 0.1273 \\
 & \cellcolor[HTML]{96FFFB}SpecTran & \cellcolor[HTML]{96FFFB}\textbf{0.0369} & \cellcolor[HTML]{96FFFB}\textbf{0.0464} & \cellcolor[HTML]{96FFFB}\textbf{0.0796} & \cellcolor[HTML]{96FFFB}\textbf{0.1173} & \cellcolor[HTML]{96FFFB}\textbf{0.0339} & \cellcolor[HTML]{96FFFB}\textbf{0.0430} & \cellcolor[HTML]{96FFFB}\textbf{0.0719} & \cellcolor[HTML]{96FFFB}\textbf{0.1084} & \cellcolor[HTML]{96FFFB}\textbf{0.0170} & \cellcolor[HTML]{96FFFB}\textbf{0.0226} & \cellcolor[HTML]{96FFFB}\textbf{0.0360} & \cellcolor[HTML]{96FFFB}\textbf{0.0582} & \cellcolor[HTML]{96FFFB}\textbf{0.0445} & \cellcolor[HTML]{96FFFB}\textbf{0.0578} & \cellcolor[HTML]{96FFFB}\textbf{0.0930} & \cellcolor[HTML]{96FFFB}\textbf{0.1463} \\ \cmidrule(l){2-18} 
\multirow{-10}{*}{HSTU} & Impr. & {\color[HTML]{C00000} \textbf{10.48\%}} & {\color[HTML]{C00000} \textbf{9.69\%}} & {\color[HTML]{C00000} \textbf{13.39\%}} & {\color[HTML]{C00000} \textbf{11.08\%}} & {\color[HTML]{C00000} \textbf{10.78\%}} & {\color[HTML]{C00000} \textbf{8.31\%}} & {\color[HTML]{C00000} \textbf{8.77\%}} & {\color[HTML]{C00000} \textbf{6.07\%}} & {\color[HTML]{C00000} \textbf{8.28\%}} & {\color[HTML]{C00000} \textbf{8.65\%}} & {\color[HTML]{C00000} \textbf{8.11\%}} & {\color[HTML]{C00000} \textbf{9.19\%}} & {\color[HTML]{C00000} \textbf{10.97\%}} & {\color[HTML]{C00000} \textbf{5.09\%}} & {\color[HTML]{C00000} \textbf{13.28\%}} & {\color[HTML]{C00000} \textbf{4.43\%}} \\
\bottomrule
\end{tabular}
}
\end{table*}

\subsection{Performance Comparison (RQ1)}
Table~\ref{tab:main} shows the performance comparison of the proposed SpecTran against the baseline methods. We observe that:

1) \textbf{Overall performance comparisons.} SpecTran significantly outperforms various sequential recommendation backbones and semantic embedding transformation baselines in terms of all evaluation metrics on four real-world datasets, and surpasses the SOTA baseline with an average improvement of 9.17\%. These results demonstrate the effectiveness of SpecTran in capturing beneficial semantic information for the recommendation model. Moreover, SpecTran can be seamlessly integrated as a plugin into different sequential recommenders to inject more comprehensive semantic knowledge for high-quality item embedding and better ranking performance.

2) \textbf{Compared with Adapter‑based baselines.}
SpecTran outperforms Adapter‑based baselines across all metrics and datasets. These Adapter‑based baselines integrate item semantic embedding into ID embeddings by different learnable MLP adapters. However, these methods have the following limitations: (i) The Adapter‑based embedding suffer from a severe collapse problem, where only a small number of dimensions contain valid information while most of the remaining dimensions are wasted.  (ii) Adapter‑based transformation struggles to distinguish the spectral weights between principal  spectral components and subordinate ones, hindering efficient extraction of semantic knowledge beneficial to traditional recommendation models.


3) \textbf{Compared with SVD-based baselines.}
SpecTran exhibits substantial improvements over the SVD-based baselines. Specifically, for WhitenRec which combines SVD and MLP for transformation, we observe that its performance is inferior to ours. It indicates that simply using the MLP  to learn spectral information is insufficient. As for LLMInit and AlphaFuse, which enhance the backbones through semantic initialization or learning ID embeddings within the null-space of semantic embeddings, their static SVD-based dimension methods lack flexibility to adapt to specific recommendation tasks. Besides, all these SVD-based baselines overlook the important subordinate spectral components. In contrast, SpecTran can effectively alleviate the challenges of static spectral weights and subordinate spectral information loss, where a Taylor expansion-based position encoding is applied for principal  spectral reweighting and a sparsified global attention mechanism is applied to aggregate both principal  and subordinate spectral components.

\begin{table*}[]
\centering
\caption{Ablation study on SpecTran.}
\label{tab:ab}
\scalebox{0.76}{
\begin{tabular}{@{}c|c|cccc|cccc|cccc|cccc@{}}
\toprule
 &  & \multicolumn{4}{c|}{\textbf{Toy}} & \multicolumn{4}{c|}{\textbf{Beauty}} & \multicolumn{4}{c|}{\textbf{Clothing}} & \multicolumn{4}{c}{\textbf{Office}} \\ \cmidrule(l){3-18} 
\multirow{-2}{*}{\textbf{Backbone}} & \multirow{-2}{*}{\textbf{Method}} & N@10 & N@20 & H@10 & H@20 & N@10 & N@20 & H@10 & H@20 & N@10 & N@20 & H@10 & H@20 & N@10 & N@20 & H@10 & H@20 \\ \midrule
 & Base & 0.0189 & 0.0247 & 0.0381 & 0.0614 & 0.0214 & 0.0286 & 0.0429 & 0.0716 & 0.0071 & 0.0098 & 0.0149 & 0.0256 & 0.0413 & 0.0549 & 0.0805 & 0.1346 \\
 & \cellcolor[HTML]{96FFFB}SpecTran & \cellcolor[HTML]{96FFFB}\textbf{0.0321} & \cellcolor[HTML]{96FFFB}\textbf{0.0415} & \cellcolor[HTML]{96FFFB}\textbf{0.0674} & \cellcolor[HTML]{96FFFB}\textbf{0.1049} & \cellcolor[HTML]{96FFFB}\textbf{0.0281} & \cellcolor[HTML]{96FFFB}\textbf{0.0381} & \cellcolor[HTML]{96FFFB}\textbf{0.0596} & \cellcolor[HTML]{96FFFB}\textbf{0.0994} & \cellcolor[HTML]{96FFFB}\textbf{0.0139} & \cellcolor[HTML]{96FFFB}\textbf{0.0193} & \cellcolor[HTML]{96FFFB}\textbf{0.0299} & \cellcolor[HTML]{96FFFB}\textbf{0.0516} & \cellcolor[HTML]{96FFFB}\textbf{0.0489} & \cellcolor[HTML]{96FFFB}\textbf{0.0629} & \cellcolor[HTML]{96FFFB}\textbf{0.0983} & \cellcolor[HTML]{96FFFB}\textbf{0.1540} \\
 & $ w/o\; \phi({\mathbf{QK}^T})$ & 0.0249 & 0.0333 & 0.0512 & 0.0848 & 0.0264 & 0.0358 & 0.0561 & 0.0932 & 0.0131 & 0.0181 & 0.0269 & 0.0471 & 0.0444 & 0.0587 & 0.0911 & 0.1481 \\
  & $ w/o\; \text{Softshrink}$ & 0.0273 & 0.0366 & 0.0567 & 0.0939 & 0.0265 & 0.0359 & 0.0558 & 0.0935 & 0.0132 & 0.0181 & 0.0267 & 0.0464 & 0.0470 & 0.0620 & 0.0956 & 0.1535 \\
 & $ w/o\; \textbf{D}_d (\rightarrow \mathbf{\Sigma}_d)$ & 0.0257 & 0.0346 & 0.0549 & 0.0900 & 0.0256 & 0.0340 & 0.0519 & 0.0854 & 0.0109 & 0.0151 & 0.0234 & 0.0398 & 0.0450 & 0.0600 & 0.0893 & 0.1491 \\
 & $w/o\; \textbf{D}_d (\rightarrow\textbf{I}_d)$ & 0.0205 & 0.0277 & 0.0433 & 0.0720 & 0.0240 & 0.0313 & 0.0497 & 0.0788 & 0.0075 & 0.0105 & 0.0159 & 0.0275 & 0.0409 & 0.0521 & 0.0787 & 0.1234 \\
\multirow{-6}{*}{BERT4Rec} & $ w/o\; \textbf{A}$ & 0.0213 & 0.0280 & 0.0451 & 0.0718 & 0.0238 & 0.0314 & 0.0482 & 0.0787 & 0.0076 & 0.0106 & 0.0157 & 0.0275 & 0.0385 & 0.0516 & 0.0737 & 0.1260 \\ \midrule
 & Base & 0.0246 & 0.0281 & 0.0482 & 0.0626 & 0.0239 & 0.0295 & 0.0496 & 0.0719 & 0.0079 & 0.0096 & 0.0161 & 0.0230 & 0.0309 & 0.0423 & 0.0674 & 0.1130 \\
 & \cellcolor[HTML]{96FFFB}SpecTran & \cellcolor[HTML]{96FFFB}\textbf{0.0390} & \cellcolor[HTML]{96FFFB}\textbf{0.0483} & \cellcolor[HTML]{96FFFB}\textbf{0.0856} & \cellcolor[HTML]{96FFFB}\textbf{0.1224} & \cellcolor[HTML]{96FFFB}\textbf{0.0340} & \cellcolor[HTML]{96FFFB}\textbf{0.0437} & \cellcolor[HTML]{96FFFB}\textbf{0.0746} & \cellcolor[HTML]{96FFFB}\textbf{0.1136} & \cellcolor[HTML]{96FFFB}\textbf{0.0175} & \cellcolor[HTML]{96FFFB}\textbf{0.0228} & \cellcolor[HTML]{96FFFB}\textbf{0.0378} & \cellcolor[HTML]{96FFFB}\textbf{0.0588} & \cellcolor[HTML]{96FFFB}\textbf{0.0471} & \cellcolor[HTML]{96FFFB}\textbf{0.0612} & \cellcolor[HTML]{96FFFB}\textbf{0.1003} & \cellcolor[HTML]{96FFFB}\textbf{0.1567} \\
 & $ w/o\; \phi({\mathbf{QK}^T})$ & 0.0385 & 0.0477 & 0.0843 & 0.1209 & 0.0327 & 0.0423 & 0.0716 & 0.1100 & 0.0169 & 0.0226 & 0.0358 & 0.0582 & 0.0464 & 0.0608 & 0.0979 & 0.1553 \\
 & $ w/o\; \text{Softshrink}$ & 0.0379 & 0.0472 & 0.0830 & 0.1196 & 0.0328 & 0.0423 & 0.0691 & 0.1071 & 0.0170 & 0.0224 & 0.0359 & 0.0576 & 0.0456 & 0.0611 & 0.0954 & 0.1566 \\
 & $ w/o\; \textbf{D}_d (\rightarrow \mathbf{\Sigma}_d)$ & 0.0366 & 0.0449 & 0.0784 & 0.1114 & 0.0309 & 0.0397 & 0.0659 & 0.1008 & 0.0162 & 0.0209 & 0.0349 & 0.0536 & 0.0435 & 0.0577 & 0.0909 & 0.1473 \\
 & $w/o\; \textbf{D}_d (\rightarrow\textbf{I}_d)$ & 0.0282 & 0.0341 & 0.0555 & 0.0792 & 0.0258 & 0.0328 & 0.0529 & 0.0806 & 0.0110 & 0.0142 & 0.0233 & 0.0360 & 0.0384 & 0.0530 & 0.0825 & 0.1403 \\
\multirow{-6}{*}{SASRec} & $ w/o\; \textbf{A}$ & 0.0267 & 0.0318 & 0.0553 & 0.0755 & 0.0259 & 0.0321 & 0.0569 & 0.0818 & 0.0107 & 0.0136 & 0.0226 & 0.0345 & 0.0328 & 0.0429 & 0.0701 & 0.1103 \\ \midrule
 & Base & 0.0257 & 0.0297 & 0.0503 & 0.0660 & 0.0251 & 0.0308 & 0.0509 & 0.0734 & 0.0077 & 0.0095 & 0.0157 & 0.0228 & 0.0326 & 0.0436 & 0.0654 & 0.1091 \\
 & \cellcolor[HTML]{96FFFB}SpecTran & \cellcolor[HTML]{96FFFB}\textbf{0.0369} & \cellcolor[HTML]{96FFFB}\textbf{0.0464} & \cellcolor[HTML]{96FFFB}\textbf{0.0796} & \cellcolor[HTML]{96FFFB}\textbf{0.1173} & \cellcolor[HTML]{96FFFB}\textbf{0.0339} & \cellcolor[HTML]{96FFFB}\textbf{0.0430} & \cellcolor[HTML]{96FFFB}\textbf{0.0719} & \cellcolor[HTML]{96FFFB}\textbf{0.1084} & \cellcolor[HTML]{96FFFB}\textbf{0.0170} & \cellcolor[HTML]{96FFFB}\textbf{0.0226} & \cellcolor[HTML]{96FFFB}\textbf{0.0360} & \cellcolor[HTML]{96FFFB}\textbf{0.0582} & \cellcolor[HTML]{96FFFB}\textbf{0.0445} & \cellcolor[HTML]{96FFFB}\textbf{0.0578} & \cellcolor[HTML]{96FFFB}\textbf{0.0930} & \cellcolor[HTML]{96FFFB}\textbf{0.1463} \\
 & $ w/o\; \phi({\mathbf{QK}^T})$ & 0.0362 & 0.0453 & 0.0779 & 0.1141 & 0.0332 & 0.0420 & 0.0703 & 0.1052 & 0.0165 & 0.0221 & 0.0349 & 0.0569 & 0.0437 & 0.0572 & 0.0907 & 0.1448 \\
   & $ w/o\; \text{Softshrink}$ & 0.0360 & 0.0446 & 0.0790 & 0.1131 & 0.0329 & 0.0419 & 0.0702 & 0.1062 & 0.0167 & 0.0221 & 0.0359 & 0.0573 & 0.0369 & 0.0475 & 0.0848 & 0.1269 \\
 & $ w/o\; \textbf{D}_d (\rightarrow \mathbf{\Sigma}_d)$ & 0.0342 & 0.0419 & 0.0703 & 0.1008 & 0.0324 & 0.0419 & 0.0696 & 0.1072 & 0.0150 & 0.0201 & 0.0328 & 0.0528 & 0.0442 & 0.0575 & 0.0913 & 0.1442 \\
 & $w/o\; \textbf{D}_d (\rightarrow\textbf{I}_d)$ & 0.0289 & 0.0352 & 0.0593 & 0.0847 & 0.0303 & 0.0389 & 0.0644 & 0.0985 & 0.0106 & 0.0135 & 0.0229 & 0.0348 & 0.0409 & 0.0539 & 0.0850 & 0.1367 \\
\multirow{-6}{*}{HSTU} & $ w/o\; \textbf{A}$ & 0.0259 & 0.0310 & 0.0546 & 0.0751 & 0.0280 & 0.0355 & 0.0582 & 0.0882 & 0.0099 & 0.0130 & 0.0216 & 0.0338 & 0.0329 & 0.0440 & 0.0715 & 0.1158 \\ \bottomrule
\end{tabular}
}
\end{table*}

\subsection{Ablation Study (RQ2)}
As shown in Table~\ref{tab:ab}, we perform the following ablation study to investigate the effects of each component: 
1) "$ w/o\; \phi({\mathbf{QK}^T})$": We remove the entire sparsified activated global attention aggregation $\phi({\mathbf{QK}^T})$ in SpecTran.
2) "$ w/o\; \text{Softshrink}$": We replace the   activation function Softshrink with the vanilla  Softmax in SpecTran.
3) "$w/o\; \textbf{D}_d (\rightarrow \mathbf{\Sigma}_d)$" and "$w/o\; \textbf{D}_d (\rightarrow\textbf{I}_d)$": We remove the Taylor expansion-based principal  spectral reweighting module by replacing the proposed $\textbf{D}_d$ with the original singular matrix $\mathbf{\Sigma}_d$ and the identity matrix $\textbf{I}_d$. 
4) "$ w/o\; \textbf{A}$": We remove the Taylor expansion-based position encoding by replacing the proposed $\textbf{A}$ with all-zero matrix. We can observe that: 

1) Overall, removing the proposed $\phi({\mathbf{QK}^T})$ from SpecTran leads to a significant decline in recommendation performance. This demonstrates the importance of SpecTran's sparsified activated global attention aggregation strategy, which can simultaneously integrate principal  and subordinate spectral components  for improving recommendation performance. 
2) Replacing the sparsified activation function Softshrink with original Softmax leads to sub-optimal performance. This is because the learned attention weights tend to be relatively flat and uniformly distributed under softmax function, which renders the adapter ineffective as a spectral selector.
3) Replacing the $\textbf{D}_d$ with $\mathbf{\Sigma}_d$ leads to a significant performance drop. This highlights the limitations of directly using the singular values as the principal  spectral weights in the original SVD-based models: The static principal  spectral weights cannot be adaptively adjusted according to the specific dataset in the recommendation model.
4) Replacing the $\textbf{D}_d$ with $\textbf{I}_d$ leads to the most obvious performance degradation. This result further demonstrates the importance of achieving a balance between preserving the original spectral weights and alleviating the anisotropy problem. A completely uniform distribution of the weights among the principal  spectral components makes it difficult for the model to capture the differences between distinct principal  spectral components. 
5) Removing $ w/o\; \textbf{A}$ leads to an obvious decline in the model's performance. This indicates that the principal  spectral components contain the main semantic information useful for the recommendation model. Ignoring the principal  spectral components will cause the item semantic embeddings to lose most of beneficial semantic information, thereby resulting in worse performance.

\subsection{Hyperparameter Sensitivity (RQ3)}
Table~\ref{tab:different_dim} showcases the performance of the proposed SpecTran against the SOTA Adapter‑based and SVD-based baseline (\eg RLMRec and AlphaFuse) on other scales of final item embedding dimension. We observe that: 1) The proposed SpecTran outperforms the backbone models, RLMRec and AlphaFuse at the dimension of 16, 32, 64, 128 and 256. This result further demonstrates the effectiveness of our method and its generalization ability on different parameter scales. 2) As the dimension decreases, SpecTran tends to show greater improvements compared with AlphaFuse. This is because the subordinate spectral components neglected by SVD-based transformation become more pronounced, while SpecTran effectively aggregates the lost semantic information through the global spectral attention, thereby achieving superior recommendation performance.

\begin{table}[]
\centering
\caption{Performance comparison under different  transformed embedding dimensions.}
\vspace{-0.3cm}
\label{tab:different_dim}
\scalebox{0.7}{
\begin{tabular}{@{}c|c|cccc|cccc@{}}
\toprule
 &  & \multicolumn{4}{c|}{\textbf{Toy}} & \multicolumn{4}{c}{\textbf{Beauty}} \\ \cmidrule(l){3-10} 
\multirow{-2}{*}{\textbf{Dim.}} & \multirow{-2}{*}{\textbf{Method}} & N@10 & N@20 & H@10 & H@20 & N@10 & N@20 & H@10 & H@20 \\ \midrule
 & Base & 0.0236 & 0.0272 & 0.0464 & 0.0606 & 0.0237 & 0.0289 & 0.0497 & 0.0708 \\
 & RLMRec & 0.0275 & 0.0343 & 0.0587 & 0.0857 & 0.0249 & 0.0328 & 0.0519 & 0.0834 \\
 & AlphaFuse & {\ul 0.0361} & {\ul 0.0452} & {\ul 0.0755} & {\ul 0.1119} & {\ul 0.0339} & {\ul 0.0432} & {\ul 0.0691} & {\ul 0.1063} \\
\multirow{-4}{*}{256} & \cellcolor[HTML]{96FFFB}SpecTran & \cellcolor[HTML]{96FFFB}\textbf{0.0398} & \cellcolor[HTML]{96FFFB}\textbf{0.0495} & \cellcolor[HTML]{96FFFB}\textbf{0.0889} & \cellcolor[HTML]{96FFFB}\textbf{0.1276} & \cellcolor[HTML]{96FFFB}\textbf{0.0359} & \cellcolor[HTML]{96FFFB}\textbf{0.0450} & \cellcolor[HTML]{96FFFB}\textbf{0.0794} & \cellcolor[HTML]{96FFFB}\textbf{0.1154} \\ \midrule
 & Base & 0.0246 & 0.0281 & 0.0482 & 0.0626 & 0.0239 & 0.0295 & 0.0496 & 0.0719 \\
 & RLMRec & 0.0295 & 0.0377 & 0.0597 & 0.0923 & 0.0290 & 0.0377 & 0.0583 & 0.0925 \\
 & AlphaFuse & {\ul 0.0358} & {\ul 0.0444} & {\ul 0.0754} & {\ul 0.1094} & {\ul 0.0336} & {\ul 0.0428} & {\ul 0.0680} & {\ul 0.1043} \\
\multirow{-4}{*}{128} & \cellcolor[HTML]{96FFFB}SpecTran & \cellcolor[HTML]{96FFFB}\textbf{0.0390} & \cellcolor[HTML]{96FFFB}\textbf{0.0483} & \cellcolor[HTML]{96FFFB}\textbf{0.0856} & \cellcolor[HTML]{96FFFB}\textbf{0.1224} & \cellcolor[HTML]{96FFFB}\textbf{0.0340} & \cellcolor[HTML]{96FFFB}\textbf{0.0437} & \cellcolor[HTML]{96FFFB}\textbf{0.0746} & \cellcolor[HTML]{96FFFB}\textbf{0.1136} \\ \midrule
 & Base & 0.0239 & 0.0276 & 0.0462 & 0.0611 & 0.0235 & 0.0291 & 0.0453 & 0.0675 \\
 & RLMRec & 0.0266 & 0.0339 & 0.0540 & 0.0832 & 0.0264 & 0.0349 & 0.0542 & 0.0875 \\
 & AlphaFuse & {\ul 0.0313} & {\ul 0.0406} & {\ul 0.0650} & {\ul 0.1024} & {\ul 0.0295} & {\ul 0.0389} & {\ul 0.0605} & {\ul 0.0978} \\
\multirow{-4}{*}{64} & \cellcolor[HTML]{96FFFB}SpecTran & \cellcolor[HTML]{96FFFB}\textbf{0.0369} & \cellcolor[HTML]{96FFFB}\textbf{0.0451} & \cellcolor[HTML]{96FFFB}\textbf{0.0771} & \cellcolor[HTML]{96FFFB}\textbf{0.1099} & \cellcolor[HTML]{96FFFB}\textbf{0.0319} & \cellcolor[HTML]{96FFFB}\textbf{0.0411} & \cellcolor[HTML]{96FFFB}\textbf{0.0668} & \cellcolor[HTML]{96FFFB}\textbf{0.1034} \\ \midrule
 & Base & 0.0237 & 0.0275 & 0.0480 & 0.0632 & 0.0236 & 0.0293 & 0.0476 & 0.0702 \\
 & RLMRec & 0.0257 & 0.0339 & 0.0517 & {\ul 0.0845} & {\ul 0.0280} & {\ul 0.0362} & {\ul 0.0545} & {\ul 0.0871} \\
 & AlphaFuse & {\ul 0.0272} & {\ul 0.0341} & {\ul 0.0565} & 0.0842 & 0.0269 & 0.0348 & 0.0536 & 0.0853 \\
\multirow{-4}{*}{32} & \cellcolor[HTML]{96FFFB}SpecTran & \cellcolor[HTML]{96FFFB}\textbf{0.0313} & \cellcolor[HTML]{96FFFB}\textbf{0.0391} & \cellcolor[HTML]{96FFFB}\textbf{0.0637} & \cellcolor[HTML]{96FFFB}\textbf{0.0942} & \cellcolor[HTML]{96FFFB}\textbf{0.0303} & \cellcolor[HTML]{96FFFB}\textbf{0.0385} & \cellcolor[HTML]{96FFFB}\textbf{0.0610} & \cellcolor[HTML]{96FFFB}\textbf{0.0935} \\ \midrule
 & Base & 0.0199 & 0.0239 & 0.0405 & 0.0565 & 0.0229 & 0.0280 & 0.0431 & 0.0635 \\
 & RLMRec & {\ul 0.0238} & {\ul 0.0309} & {\ul 0.0478} & {\ul 0.0760} & {\ul 0.0265} & {\ul 0.0342} & {\ul 0.0537} & {\ul 0.0830} \\
 & AlphaFuse & 0.0212 & 0.0278 & 0.0421 & 0.0683 & 0.0239 & 0.0303 & 0.0459 & 0.0714 \\
\multirow{-4}{*}{16} & \cellcolor[HTML]{96FFFB}SpecTran & \cellcolor[HTML]{96FFFB}\textbf{0.0260} & \cellcolor[HTML]{96FFFB}\textbf{0.0327} & \cellcolor[HTML]{96FFFB}\textbf{0.0522} & \cellcolor[HTML]{96FFFB}\textbf{0.0792} & \cellcolor[HTML]{96FFFB}\textbf{0.0269} & \cellcolor[HTML]{96FFFB}\textbf{0.0345} & \cellcolor[HTML]{96FFFB}\textbf{0.0541} & \cellcolor[HTML]{96FFFB}\textbf{0.0839} \\ \bottomrule
\end{tabular}
}
\end{table}

\subsection{Efficiency Sensitivity (RQ4)}
Table~\ref{tab:efficiency} presents the efficiency comparison of SpecTran and baseline models on the Toy dataset with the SASRec backbone.
We report three metrics: the number of trainable parameters, the total training cost and the average inference cost.
As shown, SpecTran achieves a favorable trade-off between model size and computational efficiency. Specifically, it maintains competitive training and inference times while substantially reducing the number of trainable parameters compared to the learnable baseline models such as RLMRec and LLM-ESR.
Although SpecTran exhibits slightly higher costs compared to the backbone model SASRec, it demonstrates significantly better performance without a notable increase in computation, highlighting SpecTran's ability to generate more effective semantic embeddings in an efficient manner.


\begin{table}[]
\centering
\caption{Efficiency study on SpecTran and other baselines.}
\vspace{-0.3cm}
\label{tab:efficiency}
\scalebox{0.85}{
\begin{tabular}{@{}c|c|c|c@{}}
\toprule
\textbf{Method} & \textbf{Trainable   Parameters} & \textbf{Training Cost} & \textbf{Inference Cost} \\ \midrule
RLMRec & 9.60M & 39.10s & 0.86s \\
LLM-ESR & 9.52M & 38.41s & 0.86s \\
WhitenRec & 3.23M & 32.89s & 0.87s \\
UniSRec & 3.15M & 32.43s & 0.85s \\
\rowcolor[HTML]{96FFFB} 
SpecTran & 2.21M & 24.84s & 0.61s \\
\midrule
SASRec & 1.70M & 23.46s & 0.57s \\
 \bottomrule
\end{tabular}
}
\end{table}

\subsection{Case Study (RQ5)}
In this case study, we analyze the learned principal  and subordinate spectral weights statistics of SpecTran across four datasets on the SASRec backbone. The principal  spectral weights are obtained by adding the positional encoding to corresponding attention scores in the global attention mechanism, while the subordinate spectral weights are computed by aggregating the absolute values of all attention scores along the subordinate spectral directions. As shown in Table~\ref{tab:case}, the principal  spectral weights exhibits  lower values than the subordinate spectral weights, indicating that the subordinate spectral components indeed contain semantic information beneficial to traditional recommendation models. Although the subordinate spectral weights for individual spectral components are sparse, the cumulative effect is non-negligible, which is overlooked by the previous SVD-based methods. This further demonstrates the effectiveness of SpecTran in aggregating subordinate spectral components through the global spectral attention mechanism.


\begin{table}[]
\centering
\caption{The case study on learned principal  and subordinate spectral weights of SpecTran on different datasets.}
\vspace{-0.3cm}
\label{tab:case}
\scalebox{1}{
\begin{tabular}{@{}c|cccc@{}}
\toprule
\textbf{Spectral   Weights} &  \textbf{Toy} & \textbf{Beauty} & \textbf{Clothing} & \textbf{Office} \\ \midrule
Principal  &  36.88 & 72.00 & 63.03 & 51.60  \\ 
Subordinate &  169.20 & 527.95 & 335.91 & 52.93 \\ \bottomrule
\end{tabular}
}
\end{table}

\section{Related Work}
\subsection{Sequential Recommendation}
Sequential recommendation~\cite{xie2022contrastive, chen2018sequential, chang2021sequential, li2020time, yang2024psl,yang2025breaking,zhang2026talos} aims to predict what the users might prefer next with their historical behaviors. Existing sequential recommendation methods use sequential modeling models, such as recurrent neural networks (RNNs)~\cite{hidasi2015session}, convolutional neural networks (CNNs)~\cite{tang2018personalized} or Transformers~\cite{sun2019bert4rec,kang2018self,yang2023generic}, to model user interaction sequences. 
For instance, GRU4Rec~\cite{hidasi2015session} applies gated recurrent units to capture sequential dependencies, while Caser~\cite{tang2018personalized} employs convolutional filters to learn local interaction patterns.
Transformer-based models such as BERT4Rec~\cite{sun2019bert4rec} employs the BERT~\cite{devlin2019bert} to model user interaction sequences with Cloze-style training objective; SASRec~\cite{kang2018self} introduces causal attention mechanism to automatically learn the weights of different interaction items; and HSTU~\cite{zhai2024actions} stacks the Hierarchical Sequential Transduction Units to improve model performance on large-scale recommendation datasets.
In addition, several studies have explored how to improve the robustness of sequential recommendation models~\cite{chen2023bias,lin2025recommendation,wang2024distributionally,chen2021autodebias}.
Most current LLMs for recommendation methods also adopt the setting of sequential recommendation~\cite{hou2022towards,hu2025alphafuse}, benefiting from its ability to capture temporal dependencies while inheriting its modeling framework.

\subsection{LLMs for Recommendation}
Large Language Models (LLMs) have showcased remarkable capabilities in content comprehension and semantic reasoning~\cite{achiam2023gpt,dubey2024llama, huang2026wese}.
Recently, LLMs have sparked a surge of interest within RS~\cite{li2023prompt,qin2025d2k,zhang2023recommendation,shi2024enhancing,xu2024enhancing,wang2024cela,geng2022recommendation, cui2022m6}. There are primarily two paradigms:

\subsubsection{LLM-enhanced recommenders}
This paradigm primarily leverages the rich knowledge and reasoning capabilities of LLMs to enhance traditional recommender models~\cite{ren2024enhancing,ren2024representation,sun2024large,wang2024can,ren2023representation}.

\textbf{Semantic Embedding Enhancement.} A representative approach is to encode users/items textual information as semantic embeddings, which are then integrated into traditional recommendation models for semantic enhancement~\cite{yuan2023go,hou2022towards,ren2024representation,zhang2024id,zhang2025llminit}. 
However, due to the significant dimensional gap between the LLM semantic embeddings and the ID embeddings in traditional recommendation models (\eg $4096$ vs $64$), performing transformation on the encoded LLM semantic embeddings is necessary. Existing methods can be divided into two major categories:
(1) \textbf{Adapter‑based methods} leverage learnable adapters to perform dimensional transformation and facilitate the transition from the LLM semantic space to the collaborative space. For example, MoRec~\cite{yuan2023go} employs a dense MLP layer to perform dimensional transformation on the encoded multi-modal item embeddings. UniSRec~\cite{hou2022towards} utilizes a mixture-of-experts (MoE) adapter to enhance the expressive capacity of the MLP, generating higher-quality transformation results. RLMRec~\cite{ren2024representation} designs an alignment network between semantic embeddings and ID embeddings, where a reconstruction loss is used to guide its learning. LLM-ESR~\cite{liu2024llm} proposes a dual-perspective modeling framework that fuses semantic embeddings and collaborative embeddings through a cross-attention mechanism. 
(2) \textbf{SVD-based methods} perform transformation on semantic embeddings based on their intrinsic data characteristics by selecting the top-$d$ singular values. For example, WhitenRec~\cite{zhang2024id} applies a spectral-decomposition-based whitening transformation with a mapping adapter to the item semantic embeddings for transformation. LLMInit~\cite{zhang2025llminit} directly selects the $d$ columns with the largest variance from the LLM semantic embeddings to initialize the item ID embeddings. In contrast, AlphaFuse~\cite{hu2025alphafuse} achieves effective semantic enhancement by learning ID embeddings within the semantic-sparse subspace and employing standardization on the singular values.

Although effective, these semantic embedding enhancement methods face critical challenges: Adapter‑based methods are prone to severe dimensional collapse, leading to significant wastage of the semantic embedding dimensions; SVD-based methods struggle to mitigate the imbalance distribution of principal  singular values and overlook the potentially useful semantic information in the subordinate spectral components. In contrast, our SpecTran simultaneously alleviates the challenge of dimensional collapse while learning the principal  and subordinate spectral information in a more flexible manner.

\textbf{Other Enhancement Methods.} There are also many studies enhancing recommendation models from other perspectives. One direction is to explore the LLM's capability of data augmentation, thereby yielding higher-quality data samples and improved data robustness to noise~\cite{wei2024llmrec,wang2025llm4dsr,lin2024data,cui2025field}. For example, LLMRec~\cite{wei2024llmrec} employs LLM-based data augmentation to select user-liked and disliked items from given item candidates, thus improving the quality of data samples.
LLM4DSR~\cite{wang2025llm4dsr} investigates the capability of LLMs for data denoising by training them as denoisers and replacing potential noisy samples in user interaction sequences according to predicted probabilities.
In addition, some studies have explored leveraging knowledge distillation (KD) for LLM-enhanced recommendation, which can effectively alleviate the high inference latency caused by the large parameter scale and complex architecture of LLMs during the recommendation enhancement process~\cite{wang2024can,cui2024distillation,liu2024large}. For example, DLLM2Rec~\cite{cui2024distillation} uses knowledge distillation to capture student-friendly knowledge from LLMs for traditional recommenders enhancement. The main challenge of this paradigm lies in the significant capability and mechanism gap between LLMs and traditional recommender models, which hinders effective knowledge transfer. 
SLIM~\cite{wang2024can} distills knowledge from a huge LLM into a relatively smaller LLM, enabling the smaller model to acquire powerful semantic reasoning capabilities and be utilized to enhance traditional recommendation models.
These methods leverage the semantic reasoning capabilities of LLMs from different perspectives and effectively enhance the performance of traditional recommendation models.

\subsubsection{LLM-based recommenders} 
This paradigm directly leverages pre-trained LLMs as the backbone for recommendations~\cite{li2023e4srec,hou2023learning,zheng2024adapting,jiang2024item,tan2024llmrecsys,chen2024softmax, chen2024hllm,shi2024large,zhang2025reinforced}. Early studies explored the capabilities of LLMs in zero-shot scenarios by framing the recommendation task as language prompts~\cite{wang2023zero, liu2023first, gao2023chat, hou2024large, liu2024once}. However, due to the discrepancy between the recommendation task and the pre-training language modeling task of LLMs, these methods often perform poorly.
Subsequently, the researchers explored finetuning LLMs with recommendation data to minimize the semantic gap between recommendation and natural language modeling~\cite{bao2023tallrec,lin2024bridging,bao2025bi,liao2024llara,xu2025slmrec,kim2024large,cui2025hatllm,yang2026bear}. In addition, several studies have explored how to improve the reasoning efficiency of LLM-based recommenders~\cite{geng2024breaking,lin2024rella}. This is not the focus of our research and readers may refer to the excellent survey~\cite{liu2024large,zhao2024recommender} for more details.

\section{Conclusion}
In this paper, we reveal that current LLM semantic embedding transformation methods face significant limitations: the Adapter‑based methods suffer from severe dimension collapse while the SVD-based methods  overlook the useful semantic information in the subordinate spectrum and rely on the static or hand-crafted spectral weights. To overcome these challenges, we propose SpecTran, a novel Transformer-based spectral-aware adapter that leverages a transformer to adaptively select and fuse important spectral information, mitigating dimension collapse and enabling more comprehensive spectral exploitation.
Extensive experiments on four real-world datasets demonstrate the effectiveness of our method. In the future, it would be interesting to explore more advanced spectral attention aggregation methods that could generate embeddings with higher quality.

\begin{acks}
This work is supported by the Starry Night Science Fund of Zhejiang University Shanghai Institute for Advanced Study (SN-ZJU-SIAS-001) and the National Natural Science Foundation of China (62476244,62372399). This work is also funded by ZJU-China Unicom  Digital Security Joint Laboratory.
\end{acks}


\bibliographystyle{ACM-Reference-Format}
\bibliography{sample-base}


\end{document}